\documentclass[reprint, amsmath,amssymb, aps,prl]{revtex4-2}
\usepackage{stackengine}
\usepackage{booktabs}
\usepackage{float}

\usepackage[subrefformat=parens,labelformat=parens,caption=false]{subfig}
\usepackage{graphicx}
\usepackage{bm}
\usepackage{hyperref}
\usepackage{natbib}
\usepackage[usenames,dvipsnames]{color}
\usepackage[usenames,dvipsnames,svgnames,table]{xcolor}
\usepackage{physics}
\usepackage{soul}
\usepackage{ulem}

\usepackage{comment}
\bibpunct{\color{blue}[}{\color{blue}]}{,}{n}{}{;}
\hypersetup{
  colorlinks,
  citecolor=blue,
  linkcolor=blue,
  urlcolor=blue
  }

\newlength{\mysize}

\begin{document}
\preprint{APS/123-QED}
\title{Trivial Andreev band mimicking topological bulk gap reopening in\\ the non-local conductance of long Rashba nanowires}
\author{Richard Hess}
\author{Henry F. Legg}
\author{Daniel Loss}
\author{Jelena Klinovaja}%
\affiliation{%
 Department of Physics, University of Basel, Klingelbergstrasse 82, CH-4056 Basel, Switzerland }%
 \date{\today}
 \begin{abstract}
We consider a one-dimensional Rashba nanowire in which multiple Andreev bound states in the bulk of the nanowire form an Andreev band. We show that, under certain circumstances, this trivial Andreev band can produce an apparent closing and reopening signature of the bulk band gap in the non-local conductance of the nanowire. Furthermore, we show that the existence of the trivial bulk reopening signature (BRS) in non-local conductance is essentially unaffected by the additional presence of trivial zero-bias peaks (ZBPs) in the local conductance at either end of the nanowire. The simultaneous occurrence of a trivial BRS and ZBPs mimics the basic features required to pass the so-called `topological gap protocol'. Our results therefore provide a topologically trivial minimal model by which the applicability of this protocol can be benchmarked.
\end{abstract}

\maketitle

Majorana bound states (MBSs) are predicted to appear in the cores of vortices or at boundaries of topological superconductors \cite{kitaev2001unpaired,Fujimoto2008Topological,OregHelical2010,LutchynMajorana2010,
Sato2010NonAbelian}. The non-Abelian statistics of MBSs make them highly promising candidates for fault tolerant topological quantum computing 
\cite{Volovik1999Fermion,Read2000Paired, Ivanov2001NonAbelian,Nayak2008Non,alicea2012new,Laubscher2021Majorana}. However, so far, despite significant efforts, there has been no conclusive experimental observation of MBSs. The most heavily investigated platform purported to host MBSs are hybrid semiconductor-superconductor devices. These devices consist of a semiconductor nanowire with strong Rashba spin-orbit interaction (SOI) -- for instance InSb or InAs -- that has been brought into proximity with a superconductor -- for instance NbTiN or Al  \cite{MourikSignatures2012,Deng2012Anomalous,Das2012Zero,Churchill2013Superconductor,yu2020non}. Although the presence of zero-bias peaks (ZBPs) in local conductance measurements initially appeared promising evidence for MBSs in such devices 
  \cite{Sasaki2011Topological, MourikSignatures2012, Deng2012Anomalous,Das2012Zero, Churchill2013Superconductor},
   it was subsequently realized that the same signature could be produced by trivial effects, for instance Andreev bound states (ABSs)
\cite{Andreev1964Thermal,Andreev1966Electron,kells2012Near,Lee2012Zero, Cayao2015SNS,Ptok2017Controlling,Liu2017Andreev, reeg2018zero, Penaranda2018Quantifying,moore2018two, Vuik2019Reproducing, Woods2019Zero, Liu2019Conductance,Chen2019Ubiquitous, Awoga2019Supercurrent,Alspaugh2020Volkov, Juenger2020Magnetic, valentini2020nontopological, Prada2020From, Hess2021Local, Roshni2022Conductance}. 

Trivial mechanisms that mimic the expected experimental signatures of the topological superconducting phase have significantly complicated the search for MBSs. Several auxiliary features in the local conductance were suggested to provide further clarity for the origin of a ZBP. Examples include, oscillations around zero energy due to the overlap of the MBSs in short nanowires \cite{Prada2012Transport,sarma2012splitting,Rainis2013Towards, Dmytruk2018Suppresion, Fleckenstein2018Decaying}, the flip of  the lowest band spin polarization \cite{Szumniak2017Spin,Chevallier2018Topological},
correlated ZBPs at either end of the nanowire, and a quantized conductance peak with height $2e^2/h$ \cite{law2009majorana, akhmerov2009electrically,Flensberg2010Tunneling,Wimmer2011Quantum,Chevallier2016Tomography}. Although oscillations and correlated ZBPs have been experimentally observed \cite{Sasaki2011Topological,MourikSignatures2012,Deng2012Anomalous, Das2012Zero, Churchill2013Superconductor,AlbrechtExponential2016,Wang2022Parametric}, all these local conductance signatures can be explained by trivial mechanisms \cite{cao2019decays,Vuik2019Reproducing, pan2020generic}.

\begin{figure}[t]
\subfloat{\label{figThreeTerminalDevice}\stackinset{l}{3.1in}{t}{0.1in}{\color{white}(a)}{\stackinset{l}{3.1in}{t}{1.22in}{\color{black}(b)}{\includegraphics[width=1\columnwidth]{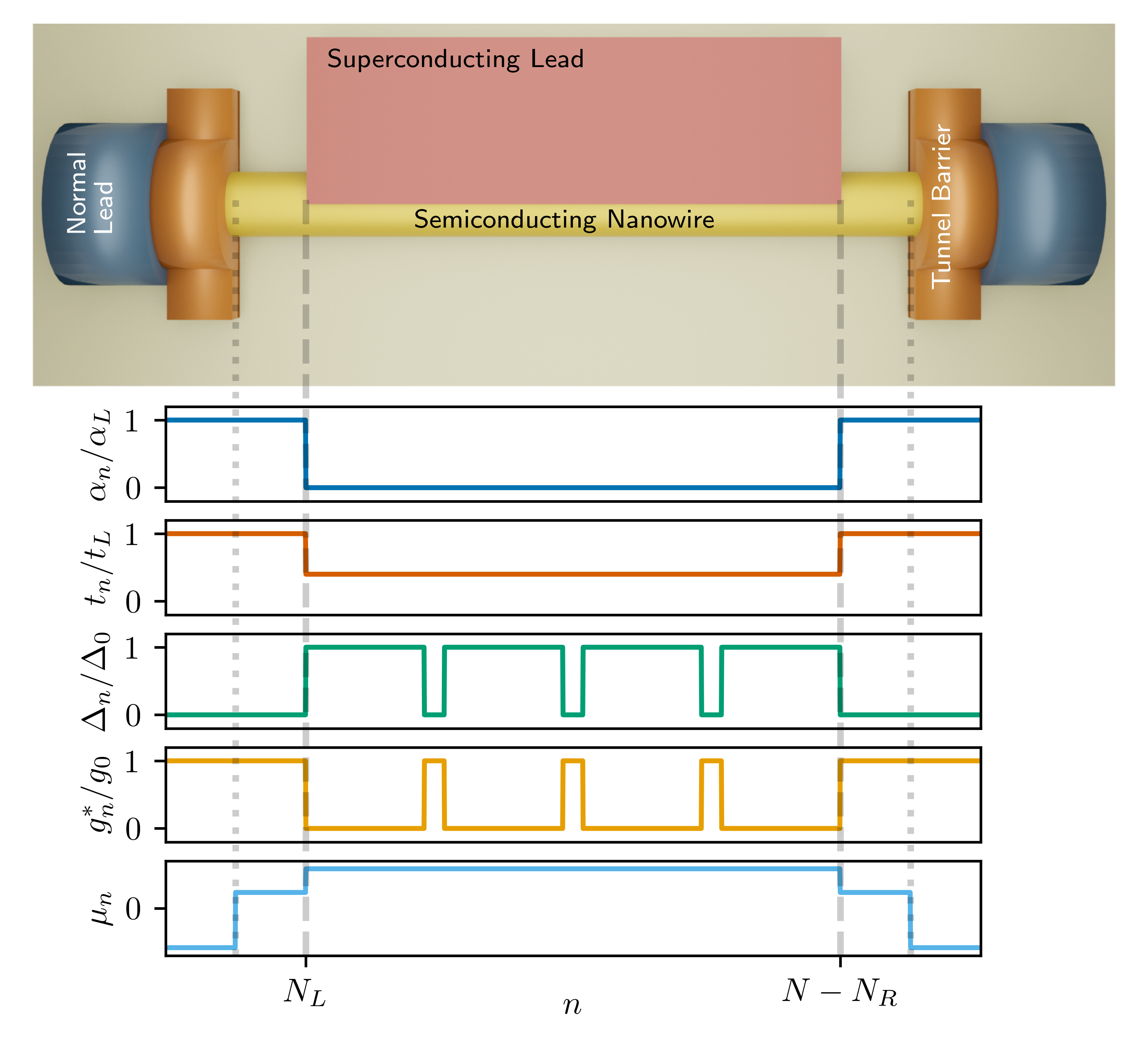}}}
} 
\subfloat{\label{figGeneralParameterProfiles}}
 \caption{ 
\textit{Schematic sketch of a three-terminal device and a typical set of parameter profiles supporting the formation of an Andreev band.} (a) A grounded superconducting lead (red) is attached to a semiconducting nanowire (yellow). Normal leads (blue), connected to the ends of the nanowire and tunnel barriers (orange), control the transparency of the interface between the normal leads and the nanowire. Experimentally several different device architectures exist but the basic features for theoretical modeling remain the same in all cases. (b) Typical parameter profile used to model equally spatially distributed ABSs and ZBPs at either end of the nanowire.}
 \label{figSchematicsDeviceAndMechanism}
\end{figure}

Subsequently, it was proposed that non-local conductance measurements in three-terminal devices -- for instance,  as shown in Fig.~\subref*{figThreeTerminalDevice} -- can detect the bulk gap closing and reopening that is associated with the phase transition to topological superconductivity, potentially providing a signature for the bulk topology of the nanowire \cite{Entin2008Conductance,LobosTunneling2014,Gramich2017Andreev,
Rosdahl2018Andreev,Zhang2019Next,DanonNonlocal2020,Melo2021Conductance, Pan2021Three,pikulin2021protocol,Banerjee2022Local}. In particular, it is important that the length of the proximitized region in a device is much longer than the  localization length of the induced superconductivity in the nanowire, otherwise  a trivial bulk reopening signature (BRS) can arise simply due to the avoided crossing of close to zero-energy ABSs \cite{Hess2021Local}. When arising due to a topological phase transition, the BRS in non-local conductance provides an upper bound for the size of the topological energy gap~\cite{pikulin2021protocol,Microsoft2022Passing}.

Based on these ideas, a so-called {\it topological gap protocol} has been proposed \cite{pikulin2021protocol}. The basic features required to pass this protocol are correlated ZBPs at either end of the nanowire in combination with a BRS. Recently, state-of-the-art experimental devices were reported to have passed this protocol \cite{Microsoft2022Passing}. 

In this paper we consider trivial mechanisms that can mimic the basic features of the topological gap protocol in nanowire devices, where the length of the proximitized nanowire is significantly longer than the localization length of the induced superconductivity. While trivial origins of ZBPs have been discussed extensively in the literature \cite{kells2012Near,Lee2012Zero, Cayao2015SNS,Ptok2017Controlling,Liu2017Andreev, reeg2018zero, Penaranda2018Quantifying,moore2018two, Vuik2019Reproducing, Woods2019Zero, Liu2019Conductance,Chen2019Ubiquitous, Prada2020From,Alspaugh2020Volkov, Juenger2020Magnetic, valentini2020nontopological, Hess2021Local, Roshni2022Conductance}, trivial mechanisms that mimic the BRS are much less understood. First, we show that it is possible for multiple ABSs to form a band inside the superconducting gap. In particular, when approximately periodically spatially distributed and at similar energies, the states within the Andreev band can have a finite support throughout the nanowire. We find that, due to the extended nature of the states in the Andreev band, they can result in a non-local conductance signal reminiscent of a BRS. Furthermore, we combine this trivial BRS with known mechanisms for trivial ZBPs at each end of the nanowire and show that a trivial BRS and correlated ZBPs can occur independently. Finally we discuss the consequences for future experimental probes of the topological superconducting phase in nanowire devices.

{\it Model:} The real-space Hamiltonian of the one-dimensional Rashba nanowire, brought into proximity with a superconductor and subject to an external magnetic field, is given by \cite{OregHelical2010,LutchynMajorana2010,Hess2021Local}
\begin{align}
H=&\sum_{n=1}^{N} \Bigg( \sum_{\nu,\nu'}c_{n,\nu}^{\dagger}\bigg[ \left( -t_{n+\frac{1}{2}}\delta_{\nu \nu'}+i\alpha_{n+\frac{1}{2}}\sigma_{\nu \nu'}^z\right) c_{n+1,\nu'}\nonumber\\
&+\frac{1}{2}\bigg\{\big(t_{n+\frac{1}{2}}+t_{n-\frac{1}{2}}-\mu_n\big)\delta_{\nu \nu'} 
+\Delta_{Z,n}\sigma_{\nu \nu'}^x \bigg\} c_{n,\nu'}\bigg] \nonumber \\
&+\Delta_n  c_{n,\downarrow}^{\dagger}c_{n,\uparrow}^{\dagger}  +\text{H.c.} \Bigg) \label{eq:effHam},
\end{align}
where $c_{n,\nu}^{\dagger}\, (c_{n,\nu})$ creates (annihilates) an electron at site $n$ with spin $\nu=\uparrow,\downarrow$ in a one-dimensional chain with a total number of $N$ sites. The Pauli matrices $\sigma^l_{\nu\nu'}$, with $l \in \lbrace x,y,z \rbrace$, act in spin space. All parameter profiles, namely hopping $t_{n\pm \frac{1}{2}}$, the proximity induced superconducting gap $\Delta_n$, the chemical potential $\mu_n$, the Rashba SOI strength $ \alpha_n $, and the Zeeman energy $\Delta_{Z,n}$ are assumed to be position dependent, indicated by the index $n$. A typical parameter profile is shown in Fig.~\subref*{figGeneralParameterProfiles} and the full mathematical expressions used can be found in the Supplemental Material (SM) \cite{CommentSupMat}.
 
Throughout we will distinguish between {\it interior} and \textit{exterior} ABSs depending on whether a given ABS occurs in the bulk or at the ends of the nanowire, respectively. The position distinguishing the bulk and ends of the nanowires is indicated by gray dashed lines in Fig.~\subref*{figGeneralParameterProfiles}. Interior ABSs arise due to {\it interior normal sections} that are modeled by a vanishing local proximity gap and increase in $g$-factor at certain positions within the bulk of the nanowire. In particular, throughout we will consider either a periodic or almost periodic distribution of these interior normal sections over the full length of the nanowire; this will allow the formation of an Andreev band within the superconducting gap (see below). For simplicity we always create ABSs using normal sections, but a modification of $g$-factor alone is sufficient to create the Andreev band 
that results in a trivial BRS (see SM \cite{CommentSupMat}). Separately, in order to enable the nanowire to host zero-energy  exterior ABSs at its ends we also model normal sections on the left and right end, which we call  \textit{exterior normal sections}, consisting of $N_L$ and $N_R$ sites, respectively.

{\it The Andreev Band:} We first develop a mechanism for the formation of a trivial band formed from Andreev bound states inside the superconducting gap based on the interplay of multiple ABSs. We will later consider the impact of this trivial band on non-local transport and trivial ZBPs. If individual ABSs are distributed in a quasi-periodic way and if, in addition, the separation between the ABSs is of the order of the superconducting coherence length, then the individual ABSs partially overlap and hybridize to form a band of Andreev states. In contrast to the individual ABSs, which are well localized, the states within this band can have a finite support throughout the nanowire. As such, we will call this band of extended Andreev states an {\it Andreev band} due its strong similarity with the well studied Shiba band, which emerges due to overlapping Yu-Shiba-Rusinov (YSR) states in spin chains on the surface of a superconductor~\cite{Yu1965YSR,Shiba1968Classical,Rusinov1969On}. We emphasize that, unlike previous proposals for topological phases due to inhomogeneous superconductivity \cite{Hoffman2016Topological,Yoav2017Realizing}, the system we consider here remains trivial for all values of magnetic field and hence all features are entirely non-topological. In particular, this is ensured by the fact(s) that the bulk $g$-factor is zero apart from in the normal sections that form the Andreev band and/or we always ensure the SOI vanishes in the bulk of the nanowire.

Since the states in the Andreev band are extended they can connect the left and right normal lead and hence these states are visible in non-local differential conductances, which are typically measured in three-terminal devices with a setup as sketched in Fig.~\subref*{figThreeTerminalDevice}. The Andreev band emerges around the energy of the individual ABSs that form it and its band width is determined by the overlap of the ABSs, which is related to the separation length between the ABSs. We note that,  within our minimal model, the band width of the Andreev band is normally smaller than the size of the bulk superconducting gap which means that there is normally a finite gap between bulk superconducting states and Andreev band states.

 \begin{figure}[t]
\subfloat{\label{figDeltaAnd_GFactor}\stackinset{l}{-0.00in}{t}{-0.0in}{(a)}{\stackinset{l}{1.7in}{t}{-0.1in}{(b)}{\stackinset{l}{-0.00in}{t}{0.95in}{(c)}{\stackinset{l}{1.7in}{t}{0.95in}{(d)}{\includegraphics[width=1\columnwidth]{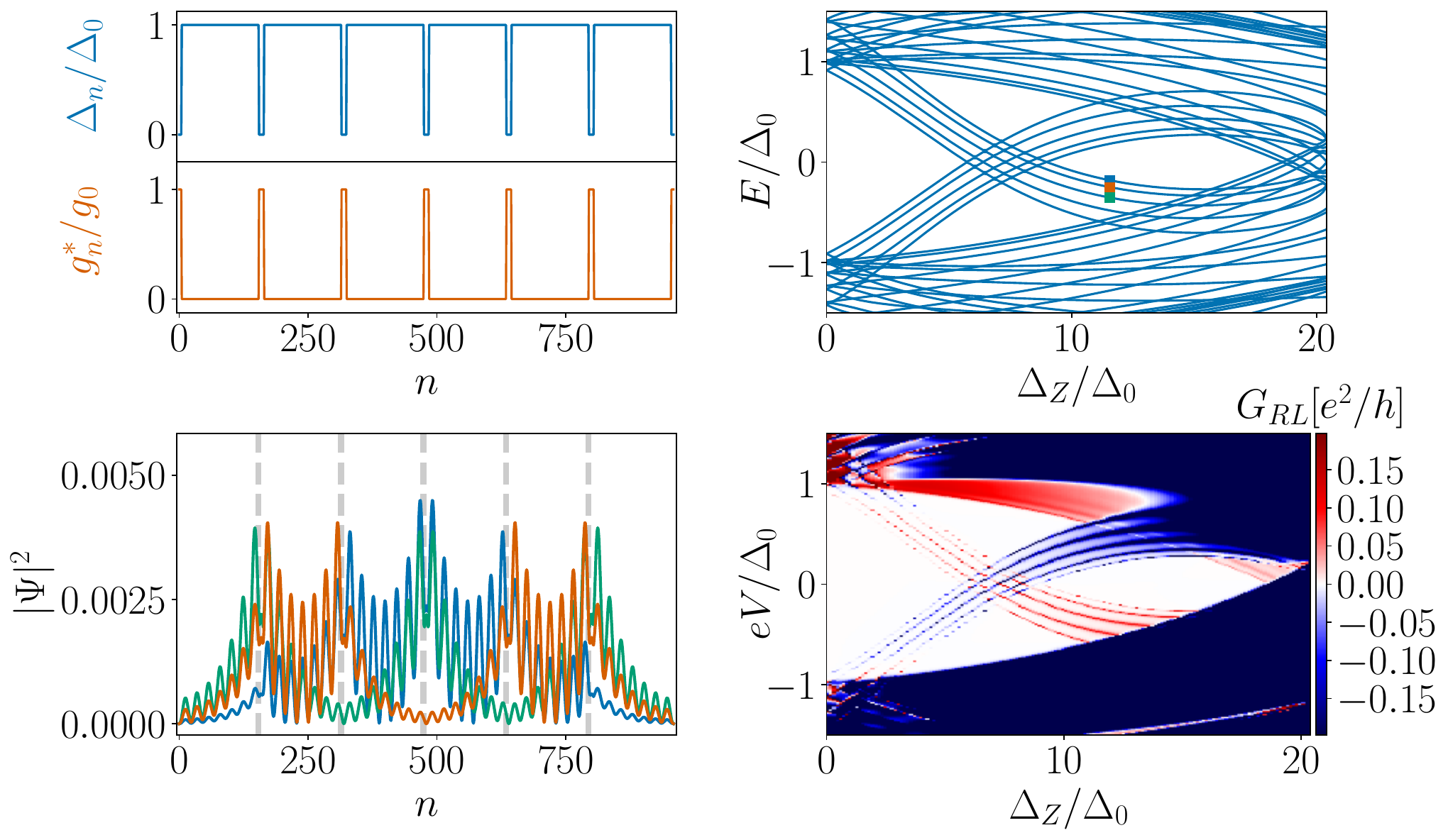}}}}}
}
\subfloat{\label{figEnergySpectrumAndreevBand}}
\subfloat{\label{figProbDensAndreevBand}}
\subfloat{\label{figNonLocalConductanceAndreevBand}}
 \caption{ 
\textit{Formation of the Andreev band}: (a) Spatial profiles of the induced superconducting gap and of the $g$-factor at zero Zeeman field in terms of their maximal zero-field values $\Delta_0$ and $g_0$, respectively. The combination of the two profiles leads to the formation of an Andreev band inside the superconducting gap. (b)  Energy spectrum of the Rashba nanowire, with Andreev band, as a function of Zeeman field. (c) Probability densities of the extended trivial Andreev states at the positions marked by the colored squares in panel (b). Gray dashed vertical  lines indicate the position of the normal sections. Due to the hybridization of ABSs that form in the normal sections, the Andreev states  extend throughout the nanowire. (d) Non-local differential conductance $G_{RL}$ as a function of Zeeman field.
Parameters:  $a=4$ nm; $(N_L=N_R=N_{B,L}=N_{B,R},N_S,N_N)=(5,150,10)$; $M=6$; $(t_L=t_R=t_{SN},\mu_L=\mu_R=\mu_{SN},\Delta_0,\Delta_Z^c,\alpha_L=\alpha_R,\gamma_L=\gamma_R, \mu_{Lead,L}= \mu_{Lead,R})\approx(158,3,0.6,12.2,0,5,5)$ meV; $T=0$. See also SM~\cite{CommentSupMat}. 
}
 \label{figAndreevBand1}
\end{figure}

{\it Trivial bulk reopening signature:} To study the transport consequences of the delocalized states in the Andreev band, we first consider a profile with periodically  distributed variations in the induced proximity gap and $g$-factor along the nanowire, as shown in Fig.~\subref*{figDeltaAnd_GFactor}. ABSs are created as a result of this profile and have the majority of their weight in the periodic normal sections. Further, these ABSs hybridize to form highly extended states, as shown by the probability densities in Fig.~\subref*{figProbDensAndreevBand}. We note that, even though we do not consider zero-energy exterior ABSs here, short exterior normal sections are present in the model to provide tunnel barriers for the differential conductance computation, which we perform with the Python package Kwant \cite{Groth2014Kwant}.

Due to the variation in the $g$-factor between the normal and superconducting sections, the energies of states which form the Andreev band have a different slope as a function of Zeeman field than states with the majority of their weight in the superconducting sections, as shown in Fig.~\subref*{figEnergySpectrumAndreevBand}. Here, the Zeeman field is defined as $\Delta_Z= g_0\mu_B B/2$, where $\mu_B$ is the Bohr magneton, $B$ the magnetic field, and $g_0$ is the $g$-factor in the normal sections. Importantly, the larger $g$-factor in the normal sections means that the Andreev band states cross zero energy considerably before the closing of the bulk superconducting gap and therefore mimic a topological  BRS in the energy spectrum. We note that for our model, the slope of the energies as a function of the Zeeman field is non-linear since we consider the superconducting gap to be a function of the Zeeman field and, in addition, the states leak into the regions with reduced $g$-factor, which means that the average Zeeman field experienced by the ABSs is reduced.

The non-local conductance only measures states that connect left and right leads and, as shown in Fig.~\subref*{figNonLocalConductanceAndreevBand}, the extended nature of the states that form the Andreev band means that it is visible in the non-local conductance. Therefore, the crossing of these states within the bulk superconducting gap mimics the bulk gap closing and reopening in non-local conductance, even though, by design, the system remains entirely trivial for all Zeeman fields.

{\it Combination of trivial effects:}
We now combine the trivial BRS due to an Andreev band with trivial ZBPs due to exterior ABSs at the ends of the nanowire, such zero-energy ABSs have previously been shown to be abundant in Rashba nanowires \cite{kells2012Near,Lee2012Zero, Cayao2015SNS,Ptok2017Controlling,Liu2017Andreev, reeg2018zero, Penaranda2018Quantifying,moore2018two, Vuik2019Reproducing, Woods2019Zero, Liu2019Conductance,Chen2019Ubiquitous, Alspaugh2020Volkov, Juenger2020Magnetic, valentini2020nontopological, Hess2021Local, Roshni2022Conductance}. Together, these trivial features mimic the key transport signatures of the topological gap protocol, namely, the exterior ABSs result in ZBPs on either end of the nanowire and the Andreev band results in a trivial BRS. To generate the ZBPs we tune the system to a certain resonance condition for SOI strength and the length of the normal sections, in order to pin the exterior ABSs to zero energy, see Refs.~\cite{reeg2018zero,Hess2021Local} for more details. However, the particular mechanism causing ZBPs at the ends of the nanowire is not the main subject of this paper and this mechanism can be exchanged for any other that results in ZBPs, as long as the formation of the Andreev band is not affected. As previously, we set the Rashba SOI strength to zero in the superconducting sections of the nanowire which ensures the system is always in the trivial phase.  In fact, the Rashba SOI is in our model only non-zero in the normal sections at the ends of the system to provide a control knob in the simulation for the zero-energy pinning of the exterior ABSs.

\begin{figure}[t]
\subfloat{\label{figEnergySpecBothAndreevPlusZBP}\stackinset{l}{-0.00in}{t}{-0.0in}{(a)}{\stackinset{l}{1.7in}{t}{0in}{(b)}{\stackinset{l}{-0.00in}{t}{1.in}{(c)}{\stackinset{l}{1.7in}{t}{1.in}{(d)}{\stackinset{l}{0.in}{t}{2.in}{(e)}{\stackinset{l}{1.7in}{t}{2.in}{(f)}{\includegraphics[width=1\columnwidth]{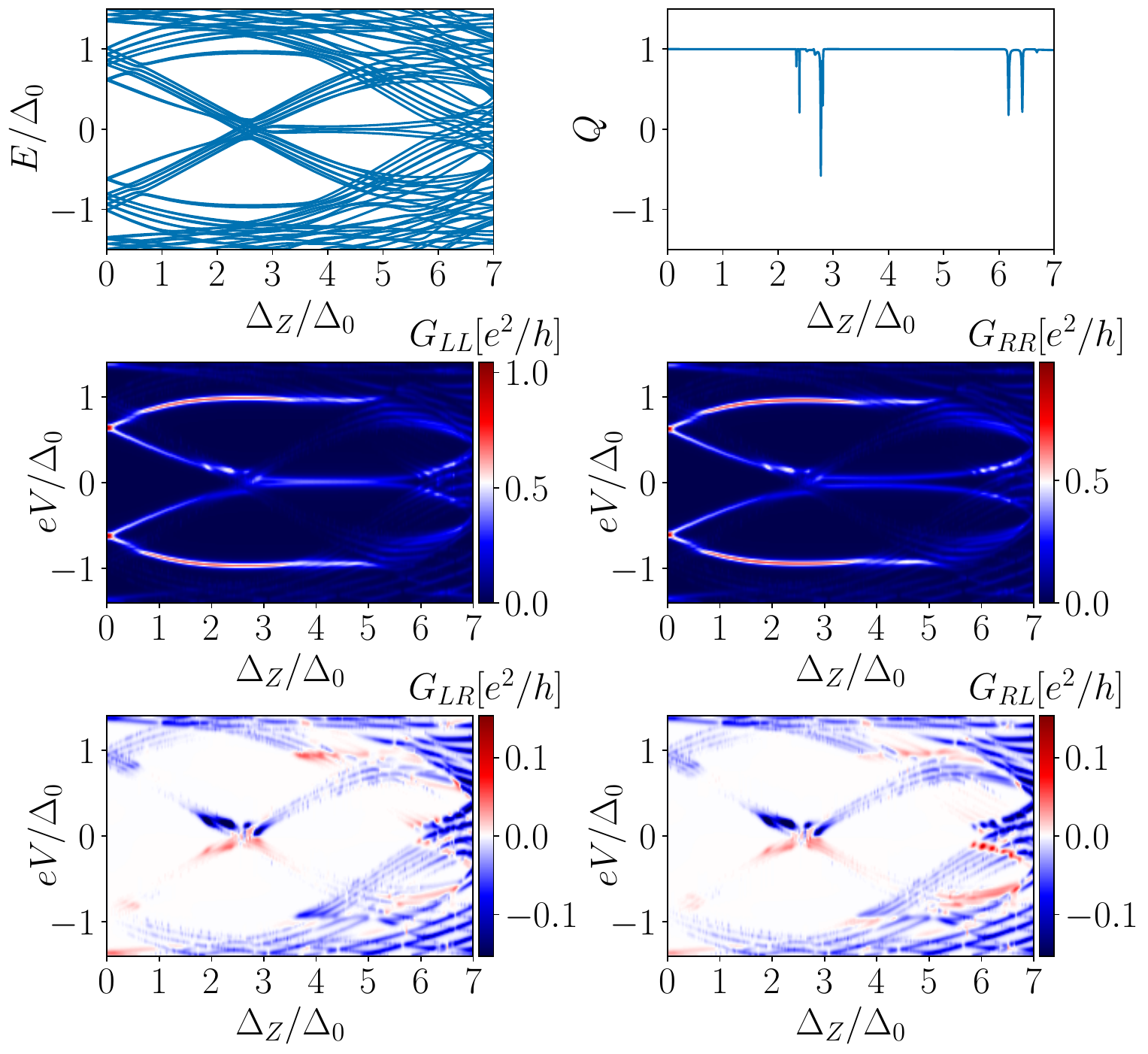}}}}}}}
}
\subfloat{\label{figQ}}
\subfloat{\label{figGLL}}
\subfloat{\label{figGRR}}
\subfloat{\label{figGLR}}
\subfloat{\label{figGRL}}
 \caption{ 
 \textit{Combination of the Andreev band with zero-energy states}: (a)  Energy spectrum showing the Andreev band crossing zero energy at the Zeeman field strength at which exterior ABSs are pinned to zero energy. (b) The topological visibility indicates the  trivial nature of the system for all Zeeman fields.  (c,d) Local differential conductance calculated on the left and on the right end of the nanowire, showing pronounced ZBPs. (e,f) Non-local conductances showing that the BRS is essentially unaffected by the presence of the ZBPs. 
Parameters: $ a=5$ nm; $(N_L,N_R,N_S,N_N,,N_{B,L}=N_{B,R})=(90,94,140,30,7)a$; $M=5$; $(t_L=t_R,t_{SN},\mu_L=\mu_R,\mu_{SN},\Delta_0,\Delta_Z^c,\alpha_L,\alpha_R,\gamma_L=\gamma_R, \mu_{Lead,L}= \mu_{Lead,R})=(100,20,0,2,0.25,1.75,13.75,13.20,10,40)$ meV; $T=40$ mK. See also SM~\cite{CommentSupMat}. 
 }
 \label{figConductanceMatrix1}
\end{figure}

In Fig.~\subref*{figEnergySpecBothAndreevPlusZBP} we show the energy spectrum of a system which combines the trivial BRS due to the Andreev band and trivial states with almost zero energy, as described above. Here, we tune the right exterior ABS slightly away from zero energy, in order to show that the exterior ABSs are independent of each other. In addition to the spectrum, we also calculate the topological visibility $Q$ \cite{Akhmerov2011Quantized,Fulga2011Scattering}, which is positive over the majority of the range of Zeeman field strengths, which is expected since our system is always in the trivial phase, see Fig.~\subref*{figQ}. We note, however, that if states of the Andreev band crosses zero energy, then the unitary property of the reflection matrix breaks down, due to the additional non-local processes, see the SM~\cite{CommentSupMat}.

Finally, the differential conductance matrix elements are shown in Figs.~\subref*{figGLL}-\subref*{figGRL}.  
We note that the ZBPs, due to the ABSs localized at both ends of the nanowire, are clearly visible in the local differential conductances, see Figs.~\subref*{figGLL} and \subref*{figGRR}, but they do not appear in the non-local conductance, as would also be the case for the ZBP signatures predicted for well separated MBSs in long nanowires \cite{Pan2021Three,Hess2021Local}. The Andreev band is visible in the non-local conductance and is also weakly pronounced in the local conductance. Hence, as previously, the Andreev band results in a conductance signature that mimics 
a topological BRS and this fact is only very weakly impacted by the addition of the trivial ZBPs at either end.

{\it Experimental relevance and outlook:}
Our mechanism for a trivial BRS requires that a few ABSs occur at similar energies and are approximately equally spatially distributed within the bulk of the nanowire. Local conductance measurements on experimental state-of-the-art Rashba nanowire devices reveal many subgap states, indicating that many ABSs remain prevalent even in high quality devices~\cite{Microsoft2022Passing}. We also note that, due to the long localization length of bound states in current nanowires (normally at least several 100~nm~\cite{Microsoft2022Passing}), the distribution only needs to include a small number of ABSs, even in nanowires with lengths of several micrometers. More precisely we find the Andreev band in systems with a length of $L\lesssim 10$ localization lengths can tolerates sizeable deviations from the periodic distribution,  see the SM \cite{CommentSupMat} for a systematic analysis. The prevalence of ABSs in current devices and the small number of ABSs required, however, means that it is likely there are regions of parameter space in current devices where the minimal scenario outlined here for the presence of a trivial BRS can occur.  Due to their independence, the probability of the combined effects occurring simultaneously, namely trivial ZBPs and trivial BRS, can be approximated from the number of occurrences of each individual signature.

We want to emphasize that the mechanism described above is much more general than the trivial BRS due to avoided crossing of ABSs in short nanowires described previously \cite{Hess2021Local}. The Andreev band mechanism for a trivial BRS is relevant for systems longer than the localization length.  Additionally, the Andreev band contains many states that contribute to the trivial BRS and is therefore much more reminiscent of a bulk signature. 

The optimal scenario for the formation of the Andreev band are ABSs with a separation very approximately equal to twice the localization length. If the separation length between the ABSs  is much larger than the localization length, then the states within the Andreev band are more localized and become only weakly visible in non-local conductance. Therefore reducing disorder, decreasing the localization length, and using longer nanowires should reduce the probability of a trivial BRS due an Andreev band. Alternatively, it might be desirable to switch device architecture or material platform to increase the phase space for MBSs and/or associated energy scales \cite{Pientka2017,Lesser2021Phase,Legg2021}.

{\it Conclusions:}
In this paper, we have shown that, when approximately equally distributed throughout the nanowire and at similar energies, multiple ABSs can hybridize to form an Andreev band within the proximity induced superconducting gap of a Rashba nanowire. This Andreev band can cross zero-energy and therefore mimics a topological bulk closing and reopening signature. The trivial BRS discussed here can be easily combined with well-known mechanisms for trivial ZBPs since we find both phenomena are essentially independent of each other.   Our simple model contains the basic features of the topological gap protocol \cite{pikulin2021protocol,Microsoft2022Passing} and thus can serve as a basis for the applicability of this protocol. In particular, such a mechanism can provide a benchmark as to whether ZBPs in local conductance and a BRS in non-local conductance alone are sufficient to distinguish a trivial and topological phase, especially when these signatures occur in only a small region of phase space.

\section{Acknowledgements}
We thank Lucas Casparis, Yuval Oreg, and Saulius Vaitiekėnas for useful conversations. This project has received funding from the European Union's Horizon 2020 research and innovation programme under Grant Agreement No 862046 and under  Grant Agreement No 757725 (the ERC Starting Grant). This work was supported by the Georg H. Endress Foundation and the Swiss National Science Foundation.

\appendix

\bibliographystyle{apsrev4-1}
\bibliography{Literatur2}
\end{document}


\preprint{APS/123-QED}
\title{Supplemental material: Trivial Andreev band mimicking topological bulk gap reopening in the non-local conductance of long Rashba nanowires}
\author{Richard Hess}
\author{Henry F. Legg}
\author{Daniel Loss}
\author{Jelena Klinovaja}%
\affiliation{%
 Department of Physics, University of Basel, Klingelbergstrasse 82, CH-4056 Basel, Switzerland }%
 \maketitle

\section{Parameter profiles}
We define the boundary between the left exterior normal section and the first section with induced superconductivity as $N_b=N_L+1/2$. If the system has in total $M$ sections each consisting of $N_{S}$ lattice sites with induced superconductivity and $M-1$ interior normal sections with $N_{N}$ lattice sites in between the superconducting sections, then we can define also a boundary between the last superconducting section, counting from left to right, and the exterior normal section on the right side as $ N_b'=N_1+(M-1)(N_S+N_N)+N_{S}+1/2$. Finally, we define the hopping matrix elements $t_n$ and the chemical potential $\mu_n$ via the auxiliary function
\begin{align}
\eta_{n}= & \eta_{L}\Theta(N_b-n)+\eta_{SN}\left[\Theta(n-N_b)-\Theta(n-N_b')\right]\nonumber +\eta_{R}\Theta(n-N_b'),
\end{align} 
where $\eta=\lbrace t,\mu\rbrace$. 
Here,  $t_{L}$ ($t_{R}$) [$\mu_{L}$ ($\mu_{R}$)] denotes the hopping matrix element [chemical potential] in the exterior normal section on the left (right) side of the nanowire, $\Theta(n)$ is the Heaviside function with the definition $\Theta(0)=1/2$, and $t_{SN}$ [$\mu_{SN}$] is the hopping matrix element [chemical potential] in the superconducting sections and the interior normal sections. We note that we do not change the hopping element (chemical potential) in the interior normal sections, since the difference $t_L-t_{SN}$ [$\mu_L-\mu_{SN}$] serves only the purpose of pinning the exterior ABSs to zero energy. 
In addition, we incorporate tunnel barriers
\begin{align}
 \gamma_n=&(\gamma_{L}+\mu_L)\Theta(N_{B}-n)+(\gamma_{R}+\mu_R)\Theta(n-N_{B}'),\label{eq:barrier}
\end{align} 
located at the ends of the nanowire, into the chemical potential via $\mu_n\rightarrow (\mu_n-\gamma_n)$.
Here $\gamma_L$ ($\gamma_R$) denotes the strength of the  left (right) tunnel barrier and we defined $N_{B}=N_{B,L}+1/2 $ ($N_{B}'=N-N_{B,R}+1/2$) for the left (right) tunnel barrier which consists of $N_{B,L}$ ($N_{B,R}$) lattice sites. We consider the SOI to be fully suppressed in the central region and to be  non-zero only in the exterior normal sections with strength $\alpha_{L}$($\alpha_{R}$) on the left (right) side, the SOI strength profile is then given by 
\begin{align}
\alpha_n=\alpha_{L}\Theta(N_b-n)+\alpha_{R}\Theta(n-N_b').
\end{align}
Additionally, we construct the profile of the induced superconducting gap as 
$\Delta_n=f_n \Delta,$
where we used the auxiliary function
\begin{align}
f_n\!=\!\!\!\sum_{j=0}^{j=M}\!\!\Lambda_n(N_L+j[ N_{S}+N_{N}],N_L+j[N_{S}+N_{N}]+ N_{S}), \label{eq:AuxFunc_f}
\end{align} which is constructed out of the rectangular function
$\Lambda_n(n_1,n_2)=\Theta( n-n_1-1/2)-\Theta(n-n_2-1/2)$.
We also make the assumption that superconductivity breaks down at a critical field strength $\Delta_Z=\Delta_Z^c$, such that 
$\Delta=\Delta_0\sqrt{1-(\Delta_Z/{\Delta_Z^c)^2}}$,
where $\Delta_0$ is the pairing potential at zero magnetic field. Consequently, the profile of the induced superconducting gap $\Delta_n$ depends on the Zeeman field strength.
Similarly to the induced superconducting gap, we define the spatially varying $g$-factor 
$g_n^*/g_0=1-f_n$,
which is suppressed in the sections with induced superconductivity, this can occur, for instance, due to the metallization effect  \cite{Reeg2018Metallization}. Here $g_0$ denotes the $g$-factor of the normal sections of the nanowire. This position dependent $g$-factor results in a Zeeman energy of the form 
$ \Delta_{Z,n}=g_n^*/g_0 \Delta_Z$,
as in the main text $\Delta_Z= g_0\mu_B B/2$ is the strength of the Zeeman field, with $\mu_B$ and $B$ denoting the Bohr magneton and the magnetic field, respectively.

\section{Minimal requirements for an Andreev band} 
\begin{figure}[t]
\subfloat{\label{figDeltaAnd_GFactor_UniformDelta}\stackinset{l}{-0.00in}{t}{-0.0in}{(a)}{\stackinset{l}{1.75in}{t}{0.0in}{(b)}{\stackinset{l}{3.5in}{t}{0in}{(c)}{\stackinset{l}{5.25in}{t}{0.in}{(d)}{\stackinset{l}{0in}{t}{1.1in}{(e)}{\stackinset{l}{1.75in}{t}{1.1in}{(f)}{\stackinset{l}{3.5in}{t}{1.1in}{(g)}{\stackinset{l}{5.25in}{t}{1.1in}{(h)}{\includegraphics[width=1\columnwidth]{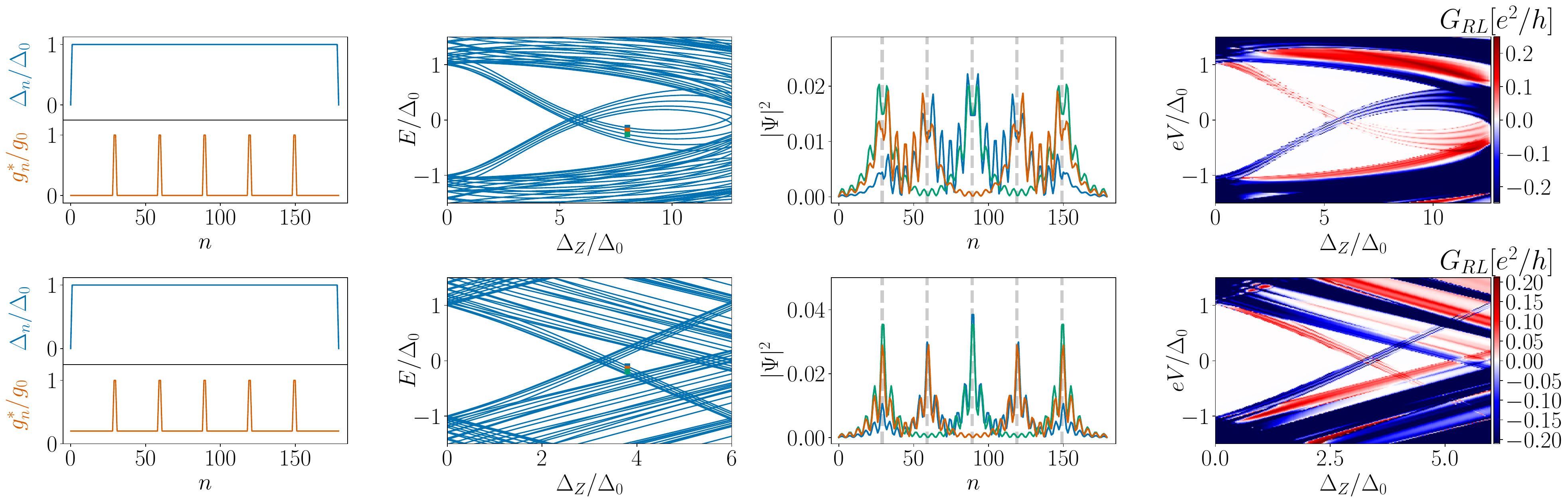}}}}}}}}}
}
\subfloat{\label{figEnergySpectrumAndreevBand_UniformDelta}}
\subfloat{\label{figProbDensAndreevBand_UniformDelta}}
\subfloat{\label{figNonLocalConductanceAndreevBand_UniformDelta}}
\subfloat{\label{figDeltaAnd_GFactor_UniformDelta_NZ_G}}
\subfloat{\label{figEnergySpectrumAndreevBand_UniformDelta_NZ_G}}
\subfloat{\label{figProbDensAndreevBand_UniformDelta_NZ_G}}
\subfloat{\label{figNonLocalConductanceAndreevBand_UniformDelta_NZ_G}}
 \caption{ 
\textit{Minimal requirements for an Andreev band}: The Andreev band appears also in case of an  uniform induced superconducting gap, normal  interior section are not required (First row). Moreover, the Andreev band forms also when the $g$-factor is non-zero in the entire nanowire, here it switches between $0.2$ and $1$ (Second row).  (a,e) Spatial profiles of the induced superconducting gap and of the $g$-factor. (b,f)  Energy spectra.  (c,g) Extended wave functions of the states marked with the coloured squares in panel (b,f). The grey dashed vertical  lines indicate the positions where the $g$-factor takes its maximum values. (d,h) Non-local differential conductance $G_{RL}$. The parameter are listed in table \ref{Tab:Parameters} and we note that we set $\Delta_Z^c\rightarrow \infty$ in the second row, which explains the linear behaviour due to the absence of SOI. The conditions on the parameter profiles, discussed in the main text, can be relaxed: The induced superconducting gap can be assumed to be constant and the $g$-factor does not need to vanish in any section of the proximitized nanowire, as long as it is modulated along the nanowire.     }
 \label{figAndreevBand1_RelaxationAndreevBandReq}
\end{figure}

In the main part of this paper, we considered a periodic distribution of interior normal sections, surrounded by superconducting sections. This structure of the induced gap was combined with an alternating $g$-factor, changing between zero and a maximal value.  Here, we consider, instead, a spatial variation of the $g$-factor combined with a {\it constant} proximity induced superconducting gap, see Fig.~\subref*{figDeltaAnd_GFactor_UniformDelta}-\subref*{figNonLocalConductanceAndreevBand_UniformDelta}. If the $g$-factor is fully suppressed in certain sections of the nanowire but non-zero in short segments or at individual spatially separated lattice sites then the model can be mapped exactly on a system containing YSR states  and consequently bound state appear for non-zero Zeeman fields \cite{Yu1965YSR,Shiba1968Classical,Rusinov1969On,Braunecker2010SpinSelective}.  In this case our suggested mechanism for the formation of a band within the superconducting gap, based on overlapping ABSs,  is equivalent to the well known Shiba band and an alternating $g$-factor alone is sufficient for the appearance of the trivial BRS. 

We note that the conditions for the Andreev band can be further relaxed: The $g$-factor can be non-zero in the whole nanowire as long as it takes much larger values at certain positions in space, as described via
\begin{align}
&g_n^*/g_0=1-g_rf_n \\ 
&\Delta_n=\Delta \Lambda_n(N_L,N-N_R), \label{eq:UniformScGap}
\end{align}
 where $g_r\in [ 0,1] $ denotes the local reduction of the $g$-factor.
 The case of $g_r=0.8$ is illustrated in Fig.~\subref*{figDeltaAnd_GFactor_UniformDelta_NZ_G}-\subref*{figNonLocalConductanceAndreevBand_UniformDelta_NZ_G}. In this case, the bulk states experience a non-zero Zeeman field and and the gap closes as a linear function of the Zeeman field strength.   The crossing point, namely the value of the Zeeman strength, at which the Andreev band states are close to zero energy, is strongly affected by the chemical potential. In particular, the crossing point moves to smaller Zeeman strengths for decreasing chemical potential, since the localization length decreases and consequently the wave function localizes around the sections with larger $g$-factor and experience therefore a stronger Zeeman field. \\

\section{Alternative Ansatz for modeling the Andreev band} 
\begin{figure}[h]
\subfloat{\label{figDeltaAnd_GFactor_JJ}\stackinset{l}{-0.00in}{t}{-0.0in}{(a)}{\stackinset{l}{1.75in}{t}{0.0in}{(b)}{\stackinset{l}{3.5in}{t}{0in}{(c)}{\stackinset{l}{5.25in}{t}{0.in}{(d)}{\includegraphics[width=1\columnwidth]{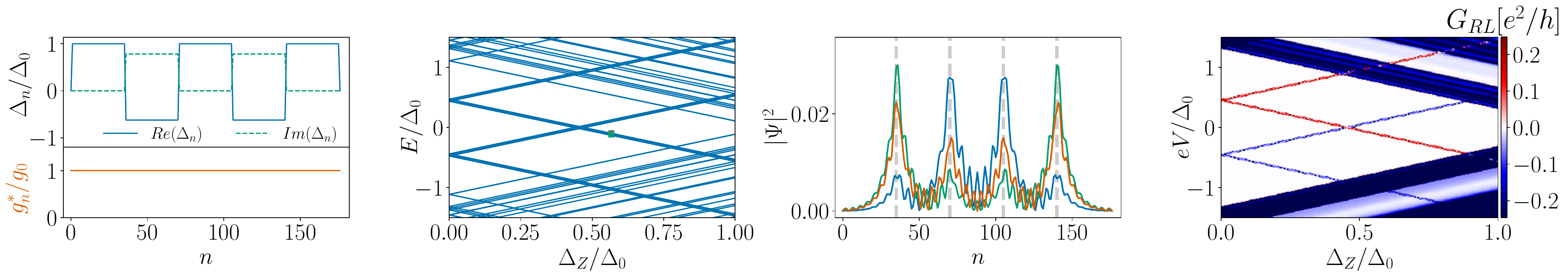}}}}}
}
\subfloat{\label{figEnergySpectrumAndreevBand_JJ}}
\subfloat{\label{figProbDensAndreevBand_JJ}}
\subfloat{\label{figNonLocalConductanceAndreevBand_JJ}}
 \caption{ 
\textit{Alternative mechanism causing the formation of the Andreev band}: We consider an induced superconducting gap with alternating phase combined with a uniform $g$-factor.  (a) Spatial profiles of the induced superconducting gap and of the $g$-factor. (b)  Energy spectra  (c) Extended wave functions of the states marked with the coloured squares in panel (b). The grey dashed vertical  lines indicate the positions where the phase of the superconductor changes (d) non-local differential conductance $G_{RL}$. The parameter are listed in table \ref{Tab:Parameters}. We conclude that the specific profiles of induced superconducting gap and $g$-factor are not relevant for the formation of the Andreev band, as long as ABSs, with approximately equal energy, are approximately periodically distributed. }
 \label{figAndreevBandDueToVaryingScPhase}
\end{figure}

Here, we note that the reason for the appearance of ABSs is actually irrelevant for the formation of the Andreev band, as long as periodically distributed ABSs with similar energy exist in the system. An alternating phase of the superconducting order parameter along the nanowire, as described via 
\begin{align}
\Delta_n=\Delta \lbrace[\Lambda_n(N_L,N-N_R)-f_n]e^{i\varphi}+f_n\rbrace, \label{eq:VaryingPhaseOfOrderParameter}
\end{align}
combined with a constant  $g$-factor  $g_n^*/g_0=1$ (see Fig.~\subref*{figDeltaAnd_GFactor_JJ}-\subref*{figNonLocalConductanceAndreevBand_JJ}), for example, also results in the appearance of ABSs in the Rashba nanowires and therefore enables the study of the formation of the Andreev band. This means that even if the ABSs in the experiment are caused by another mechanism than discussed in this paper, then the physics underlying the formation of the Andreev band and the corresponding requirements remain valid. Last we note that the parameter $N_N$ which enters Eq.~\eqref{eq:VaryingPhaseOfOrderParameter} via the auxiliary function $f_n$ denotes here the number of sites in sections with different superconducting phase.

\section{Disorder}
In this section, we  study the effect of disorder on the formation of the Andreev band. Practically, ABSs will be randomly distributed over the whole length of the nanowire. If the energies of these ABSs differ strongly, then they do not form a band. However, if multiple ABSs are present and a a subset of these ABSs are comparable in energy, then they can form a band under the condition that the remaining ABSs are sufficiently separated in space or in energy. 

Based on the above considerations we will focus on the positional variation of ABSs with equal energies.  In order to maximize the non-local conductance in the periodic case we place the first ABS approximately one localization length, $\xi$, away from the left end of the proximitized section. We note that we neglect a potential overlap of ABSs through the superconducting substrate \cite{Zu}. The distance between two adjacent ABSs is set to roughly two localization lengths and the very right ABS (the last one) is separated by one localization length from the right end of the proximitized section. The proximitized section is described via Eq.~\eqref{eq:UniformScGap}
and the critical field strength $\Delta_Z^c$ is set to infinity so that $\Delta=\Delta_0$.
All ABSs stem from the spatial variation of the $g$-factor which is modeled via  
\begin{align}
g_n=\sum_{j=1}^{j=M} \Lambda_n(n_{j,0},n_{j,0}+N_N),
\end{align} 
 where $n_{j,0}=N_L+N_S/2+jN_S+ jN_N$ denotes the left position of the $j$-th section with non-zero $g$-factor, with $j \in \lbrace 1,2,..., M\rbrace$ in case of a periodic  distribution. The disordered position can be written as  $n_{j,d}=n_{j,0}+\delta n$  where $\delta n$ is randomly chosen from a  Gaussian distributions with variance $\chi_{\delta}$ and rounded to an integer number with the condition that the non-zero $g$-factor sections lie within the superconducting section. If $N_S/2$ is not an integer, then we round the value accordingly. 
 
 In order to study the effect of positional disorder systematically in the parameter space, we compute $G_{RL}$ for different standard deviations $\chi_{\delta}$ . The larger the standard deviation the more the ABS distribution deviates from the periodic case. Note that we use the python package Adaptive \cite{Nijholt2019Adaptive} to sample the peaks of the non-local conductance with a higher resolution for this particular study.
 
In addition to the variation of the standard deviation, we calculate $G_{RL}$ for different numbers of ABSs, $M$, while keeping the density of ABSs fixed. In other words, a higher number of ABSs implies a longer nanowire. For example, we place two ABSs in a nanowire with a length of appproximately four times the localization length, while we place three ABSs in a nanowire of an approximate length of six localization lengths. We calculate $G_{RL}$ for 40 disorder configurations of each combination of $M$ and $\chi_{\delta}$, then obtain the finite temperature non-local conductance via a convolution with the derivative of the Fermi distribution. Finally, we extract the maximum of the absolute value of the finite temperature non-local conductance in the energy interval $[-0.9\Delta_0, 0.9\Delta_0]$ and average it over the 40 distributions yielding the quantity $ \langle{\rm max } \left(|G_{RL}^{intra}|\right)\rangle$. We note that the choice of the energy interval, the $g$-factor profile, and $\Delta_Z^c$ ensure that the gap does not close as a function of the Zeeman field such that we indeed measure only non-local conductance due to Andreev band states.

   \begin{figure}[t]
\subfloat{\label{figNonLocalCondDiagram}\stackinset{l}{-0.00in}{t}{-0.0in}{(a)}{\stackinset{l}{1.75in}{t}{0.0in}{(b)}{\stackinset{l}{3.5in}{t}{0in}{(c)}{\stackinset{l}{5.25in}{t}{0.in}{(d)}{\stackinset{l}{0in}{t}{1.1in}{(e)}{\stackinset{l}{1.75in}{t}{1.1in}{(f)}{\stackinset{l}{3.5in}{t}{1.1in}{(g)}{\stackinset{l}{5.25in}{t}{1.1in}{(h)}{\includegraphics[width=1\columnwidth]{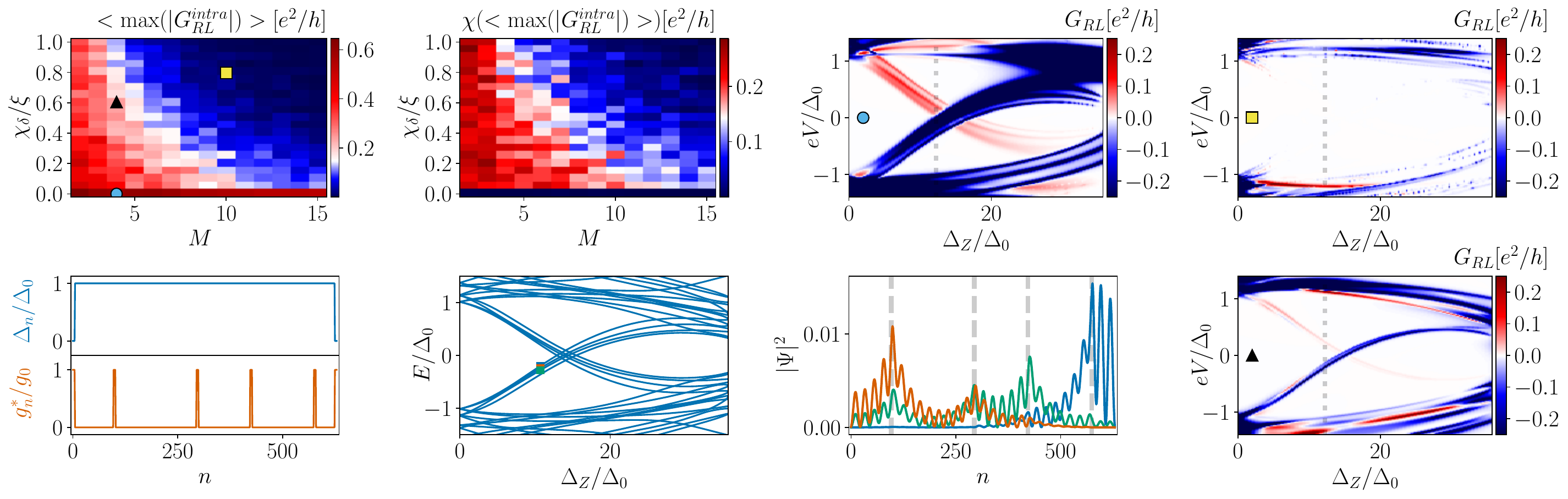}}}}}}}}}
}
\subfloat{\label{figNonLocalCondDiagram_StandDev}}
\subfloat{\label{figNonLocalCondDiagram_Periodic}}
\subfloat{\label{figNonLocalCondDiagram_Disordered}}
\subfloat{\label{figNonLocalCondDiagram_Profiles}}
\subfloat{\label{figNonLocalCondDiagram_Energies}}
\subfloat{\label{figNonLocalCondDiagram_WaveFunc}}
\subfloat{\label{figNonLocalCondDiagram_ModerateDis}}
 \caption{ 
\textit{Impact of positional disorder on the Andreev band}:   (a) Averaged maximum $\langle{\rm max } \left(|G_{RL}^{intra}|\right)\rangle$ of the absolute value of the non-local conductance in the energy interval $eV \in [-0.9\Delta,0.9\Delta]$, as a function of the standard deviation $\chi_\delta$ from the periodic distribution and of the number of ABSs $M$. We made sure that we only pick non-local conductance signatures due to the Andreev band. (b) The standard deviation of $\langle{\rm max } \left(|G_{RL}^{intra}|\right)\rangle$ from its mean value. (c,d,h) $G_{RL}$ as a function of the Zeeman field strength in case of  (c)  a periodic system, (d) a long nanowire with an ABS distribution strongly deviating from the periodic case, and (h) a nanowire with a length of eight times the localization length and with a moderate deviation from the periodic distribution. The dotted lines  indicate the Zeeman strength used in (a). (e) Spatial profiles of the induced superconducting gap and the $g$-factor illustrating the  deviation from periodicity. The configuration is associated to $G_{RL}$ shown in (h). (f) The energy spectrum and (e) probability densities of the lowest states corresponding to the system from (e) and (h). These plots show that if the nanowire is short (a few times the localization length), then the Andreev band tolerates sizeable deviation from the periodic ABS distribution and it is still visible in non-local conductance. If, in contrast, the length of the nanowire is a several orders greater than the localization length, then the wave functions are strongly localised and consequently only a distribution close to the periodic case can maintain the Andreev band and its signature in non-local conductance. The parameter are listed in table \ref{Tab:Parameters}.}
 \label{figNonLocalCondDiagramAllPlots}
\end{figure}
 
In Figs.~\ref{figNonLocalCondDiagram}  and \ref{figNonLocalCondDiagram_StandDev}, we present $ \langle{\rm max } \left(|G_{RL}^{intra}|\right)\rangle$  and the corresponding standard deviation $ \chi\langle{\rm max } \left(|G_{RL}^{intra}|\right)\rangle$. The non-local conductance is maximal in case of the periodic distribution of the ABSs and we note that the response is almost constant as a function of the number of ABSs $M$, meaning that the length of a periodic system has a rather small impact on the formation of the Andreev band and its signature in non-local conductance. If only two ABSs are present in the system, then it is not possible to define a period and consequently two ABSs of equal energy hybridize well and mediate a signal in non-local conductance signal for many positional ABS distributions, as long as the nanowire length does not exceed four times the localization length. In contrast, in much longer nanowires with more ABSs, deviations from the spatial periodic distribution of ABSs lead to a localization of the wave functions  (similar to Anderson localization in one-dimensional conductors with disorder), and therefore to a decrease of non-local conductance mediated by Andreev band states. We note, however, that in the experimental relevant regime of nanowires of a length between four and ten localization length, which corresponds to $2\leq M\leq 5$, the non-local conductance signal is still sizeable for realistic deviations from the periodic distribution. In general, we find that the longer the nanowire the less deviation from the periodic distribution is tolerated by the Andreev band signature in non-local conductance.  The standard deviation of $ \chi\left[\langle{\rm max } \left(|G_{RL}^{intra}|\right)\rangle\right]$ shown in Fig.~\ref{figNonLocalCondDiagram_StandDev} reveals that the the non-local conductance deviates much more in shorter systems and we note that the periodic system ($\chi_{\delta}=0$) has zero standard deviation, since the positions of the ABSs are fixed.

Next, we choose three different system configurations, with lengths and standard deviations specified by the markers in Fig.~\ref{figNonLocalCondDiagram}, and calculate the non-local conductance as a function of the Zeeman field strength. Here, in contrast to Figs.~\ref{figNonLocalCondDiagram}  and \ref{figNonLocalCondDiagram_StandDev}, we use a finite value of $\Delta_Z^c$ in order to close the gap at a certain Zeeman field strength and we sample the non-local conductance on an equally spaced energy grid. First, we study a nanowire with spatial periodic ABSs, see the  light blue circular  marker in Fig.~\ref{figNonLocalCondDiagram} and Fig.~\ref{figNonLocalCondDiagram_Periodic}. This distribution of ABSs supports the Andreev band, as discussed in the main text. Next, we analyse a long nanowire with ABSs deviating strongly from a periodic distribution. This deviation leads to the localization of the wave functions and non-local conductance is strongly reduced, compared to the periodic case, inside the gap independent of the values of the Zeeman field. Last, we study the experimental relevant case of a nanowire with a moderate deviation from the periodic distribution and with a length of eight times the localization length. The profiles of proximity gap and $g$-factor are shown in Fig.~\ref{figNonLocalCondDiagram_Profiles}, and the latter profile reveals the positional disorder of the ABSs.  Figure~\ref{figNonLocalCondDiagram_Energies}  shows the energies associated with the ABS configuration  and Fig.~\ref{figNonLocalCondDiagram_WaveFunc} shows the  probability densities of the three lowest energies. The wave functions clearly begin to become localized but some of them have still a small support throughout the whole nanowire, explaining the reduced but finite non-local conductance signal shown in Fig.~\ref{figNonLocalCondDiagram_ModerateDis}.

\section{Alternative trivial bulk gap reopening mechanisms}
In the main part of this paper, we discuss the Andreev band as a signature reminiscent of a gap reopening. Here for the sake of comparison we consider a considerably simpler mechanism: If the proximity effect itself is affected by the Zeeman field strength, then the  induced gap might reach a local minimum at a Zeeman field strength $\Delta_{Z,{\rm min}}$ smaller than the critical field strength $\Delta_{Z}^c$ of the break-down of superconductivity or smaller than the critical field strength $\Delta_Z^T$ of a potential topological phase transition. Although speculative, such behaviour of the induced gap could be the result of the interplay of orbital effects in the parent superconductor, inter-facial disorder between the semiconducting nanowire and the superconductor, and/or the particular device geometry. We stress that we consider here a reduction of the induced gap $\Delta(\Delta_Z)$ at $\Delta_{Z,{\rm min}}$ and not of the gap of the parent superconductor. To model this we simply assume, without any detailed justification, that the induced gap could accidental behave as 
\begin{align}
\Delta(\Delta_Z)= \Delta_0 \left[1- \kappa_1 e^{-\left(\frac{\Delta_Z-\Delta_{Z,{\rm min}}}{\kappa_2}\right)^2}\right]\label{eq:InducedGapLocalMinimum},
\end{align}
where $\kappa_1 \in [0,1]$ determines the reduction of the gap and $\kappa_2$ sets the interval of Zeeman strength over which the gap is suppressed. The exact form of $\Delta(\Delta_Z)$ is not relevant, the main requirement for the closing and reopening is the presence of local minimum as described above. 

This mechanism is, in general, independent of any ABSs which appear in the nanowire, we can therefore combine trivial zero-energy states, residing at the ends of the nanowire, with a gap reopening signature caused by a field-dependent proximity effect. Here, we consider only one trivial sub-gap state on the left end of the nanowire, however, it is, obviously, also possible to tune additionally a second trivial sub-gap state on the right end to zero-energy. 
In contrast to the main part of this paper, we model this time a nanowire which can undergo a topological phase transition since the  Rashba SOI is chosen to be non-zero in the region with induced superconducting gap.

\begin{figure}[t]
\subfloat{\label{figEnergySpecBothAndreevPlusZBP_app}\stackinset{l}{-0.00in}{t}{-0.0in}{(a)}{\stackinset{l}{1.8in}{t}{0in}{(b)}{\stackinset{l}{-0.00in}{t}{1.in}{(c)}{\stackinset{l}{1.8in}{t}{1.in}{(d)}{\stackinset{l}{0.in}{t}{2.in}{(e)}{\stackinset{l}{1.8in}{t}{2.in}{(f)}{\includegraphics[width=0.5\columnwidth]{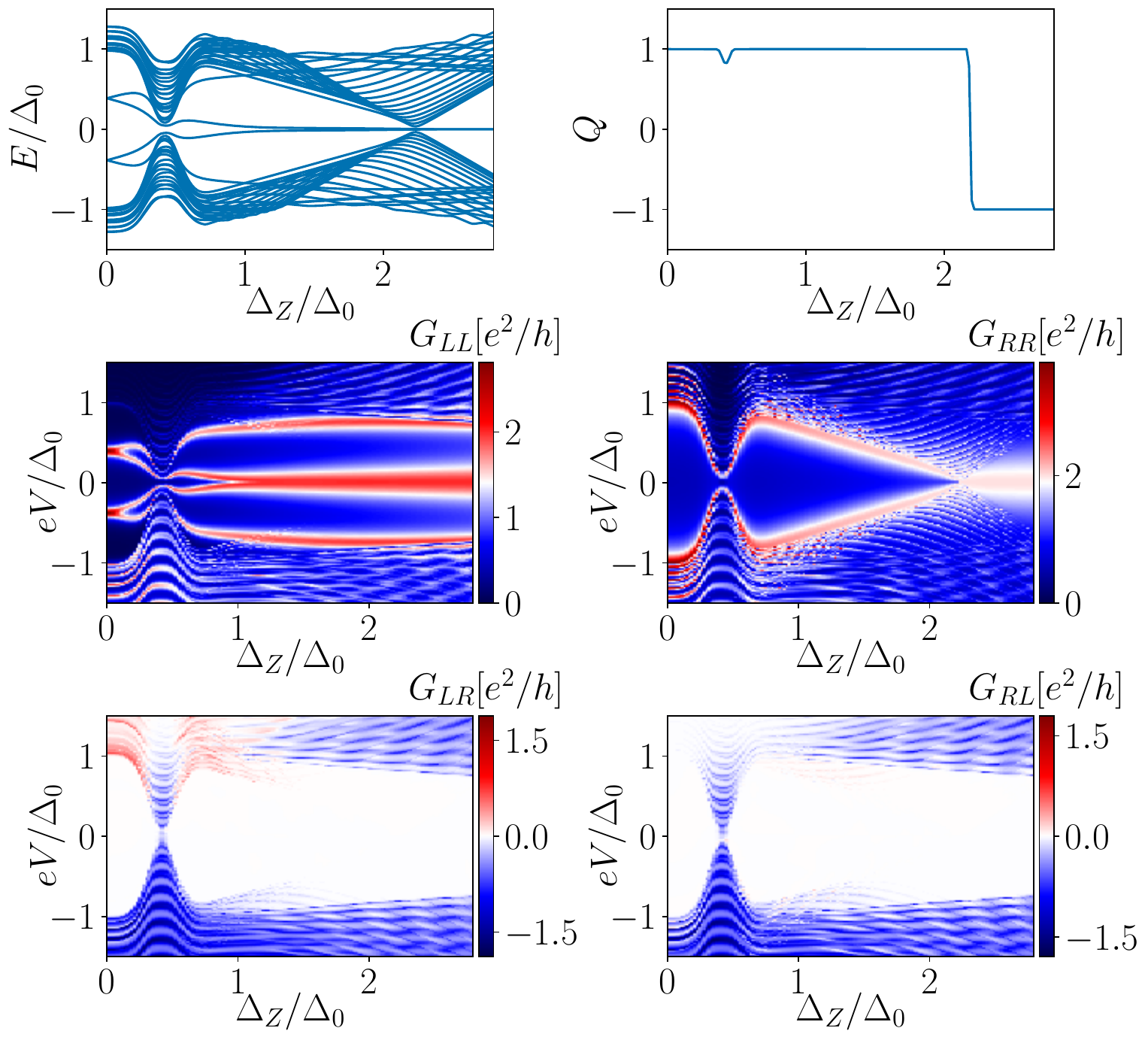}}}}}}}
}
\subfloat{\label{figQ_app}}
\subfloat{\label{figGLL_app}}
\subfloat{\label{figGRR_app}}
\subfloat{\label{figGLR_app}}
\subfloat{\label{figGRL_app}}
 \caption{ 
 \textit{Gap closing due to a Zeeman field dependent proximity effect}: (a)  Energy spectrum showing the enforced closing due to a variation of the induced superconducting gap at $\Delta_{Z,{\rm min}}=0.42 \Delta_0$ in addition to the actual topological phase transition $\Delta_Z^T\approx 2.24 \Delta_0$. (b) The topological visibility $Q$ indicates that the system remains trivial after the first gap reopening  (c,d) Local differential conductance calculated on the left and on the right end of the nanowire and (e,f) non-local conductances. The parameter are listed in Table \ref{Tab:ParametersTopologicalWire}. The trivial gap closing and reopening is visible in local and non-local conductance, while the single left ABS is only visible in left local conductance. }
 \label{figGab_app}
\end{figure}

The hopping and SOI strength profiles are simplified as $t_n=t$ and $\alpha_n=\alpha$.
 Moreover, we use a smooth step functions of the superconducting gap and chemical potential, instead of tuning the system to a resonance condition, to pin the energy of an ABS to zero for Zeeman fields smaller than the critical field 
 \begin{align}
\Delta_Z^T (\Delta_Z)=\sqrt{\left[\Delta(\Delta_Z)\right]^2+\mu_{SN}^2} \label{eq:CondTPT},
\end{align}
 associated with the topological phase transition. We note here that this condition becomes explicitly dependent on the Zeeman field strength. The profile of the superconducting gap and the chemical potential are modeled by 
\begin{align}
\Delta_n=\Delta(\Delta_Z) \begin{cases} 
\frac{1}{2}\left[1+\tanh\left(\lbrace n-N_L\rbrace/\lambda\right)\right], &  \text{if} \,\,\,  N_{B,L} <n \leq N-N_{B,R}\\
0, & \text{else} 
\end{cases}
\end{align}
and 
\begin{align}
\mu_n=\Delta(\Delta_Z) \begin{cases} 
\mu_L+(\mu_{SN}-\mu_L)\frac{1}{2}\left[1+\tanh\left(\lbrace n-N_L\rbrace/\lambda\right)\right], &  \text{if} \,\,\,  N_{B,L} <n \leq N-N_{B,R}\\
0, & \text{else},
\end{cases}
\end{align}
where $\lambda$ controls the smoothness of the step function. 
A representative energy spectrum of such a system is shown in Fig.~\subref*{figEnergySpecBothAndreevPlusZBP_app}. The gap of the system closes and reopens two times as a function of the Zeeman field: the first gap reopening is enforced via the relation defined in Eq.~\eqref{eq:InducedGapLocalMinimum}, and appears at  $\Delta_{Z,{\rm min}}$, while the second gap reopening appears at the topological phase transition and this second BRS would also emerge in case of a constant function $\Delta(\Delta_Z)=\Delta_0$. We plot the topological visibility $Q$ in Fig.~\subref*{figQ_app}. The system remains trivial between $\Delta_{Z,{\rm min}}$ and $\Delta_Z^T$. $Q$ is only negative for Zeeman strength which are approximately given by $\Delta_Z>\Delta_Z^T$. A ZBP is clearly visible in the region $\Delta_{Z,{\rm min}}<\Delta_{Z}<\Delta_Z^T$, see Fig.~\subref*{figGLL_app}. We choose a sufficiently large chemical potential so that the system remains trivial ($Q>0$) for Zeeman field strengths in the vicinity of $\Delta_{Z,{\rm min}}$.

Finally, we note that the gap closing corresponding to the topological phase transition is less pronounced in the non-local conductance than the enforced gap reopening at $\Delta_{Z,{\rm min}}$, since the associated wave functions  decay into the  normal section on the left side.

\section{Impact of the Andreev band on the topological visibility Q}

Here, we comment on the sharp negative dip of the topological visibility $Q$ presented in Fig. 3b of the main text. Originally, $Q=\text{det}[r(\omega=0)]$ was suggested as a topological index in finite size systems  of the symmetry class $D$ \cite{Akhmerov2011Quantized,Fulga2011Scattering,Brouwer2011Topological, Beenakker2011Random, Komijani2014Effect}: In two-terminal devices two processes take place if an electron, with an energy smaller than the superconducting gap, is incident on the NS interface, namely normal reflection and Andreev reflection. The reflection matrix is unitary  and particle-hole symmetric, therefore its determinant takes only the values $+1$ in case of normal reflection or  $-1$ in case of perfect Andreev reflection \cite{Akhmerov2011Quantized,Fulga2011Scattering,Brouwer2011Topological, Beenakker2011Random, Komijani2014Effect}. The latter process can be mediated by MBSs, located at the interface, due to their equal electron and hole weights \cite{law2009majorana}. Fluctuations of the barrier  strength or small disorder do not affect the unitary property of the reflection matrix and the value of the determinant does consequently not change. If the superconducting gap, however, closes, then an incident electron can enter the superconductor. This additional process leads to a break down of the unitary property of the reflection matrix and $Q$ is therefore not necessarily $\pm 1$ in gapless systems. \\
\begin{figure}[h]
\subfloat{\label{figEnergySpecBothAndreevPlusZBP_app}\stackinset{l}{-0.00in}{h}{2.4in}{(a)}{\stackinset{l}{-0.00in}{h}{1.6in}{(b)}{\includegraphics[width=0.4\columnwidth]{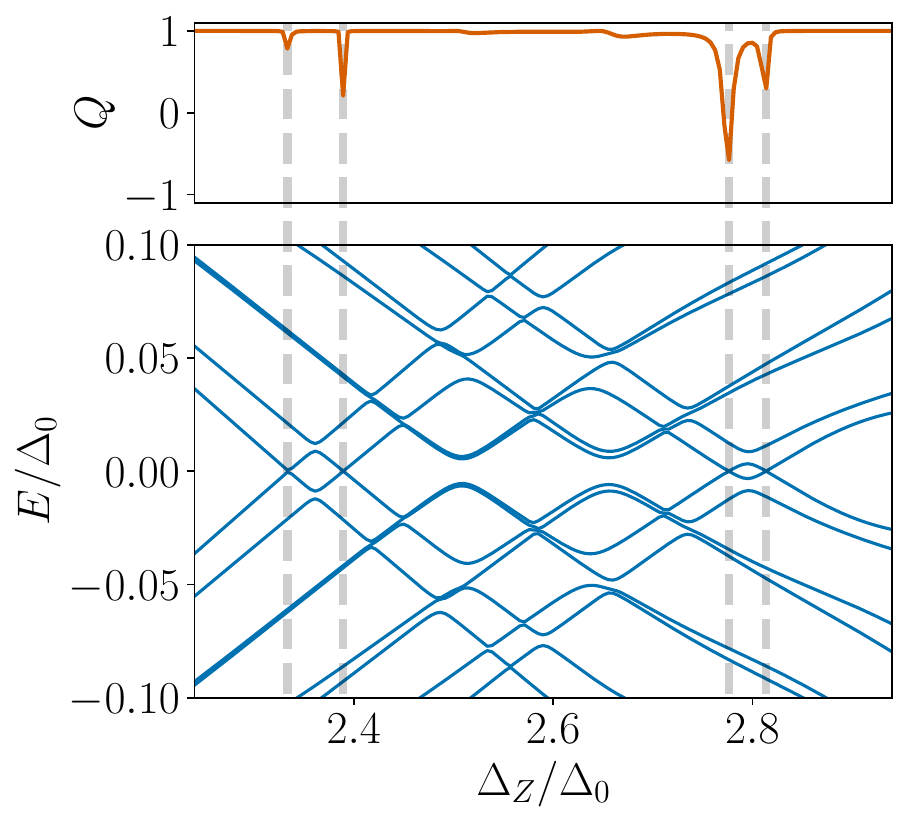}}}}
 \caption{ 
 \textit{Dips of the topological visibility $Q$}:  (a) $Q$ and (b) lowest eigenvalues as a function of the Zeeman field strength. $Q$  has dips when the lowest state of the Andreev band crosses zero-energy, this can be explained via the breakdown of the unitary property of the reflection matrix due to the additional non-local processes. The grey dashed lines serve as guides for the eye to compare the position of the dips with the roots of the lowest energy. The parameters are the same as in Fig. 3 presented in the main text.}
 \label{figGab_app}
\end{figure}
In this paper, we study a three terminal device, meaning that additional to the local processes of normal reflection and Andreev reflection also electron tunnelling  (ET) and crossed Andreev reflection (CAR) between the two normal leads take place. The Andreev band supports non-local conductance and therefore ET and CAR at certain sub-gap energies. If these processes occur at the Fermi energy, then the unitary property of the reflection matrix breaks down and enables changes of $Q$, consequently $Q$ does not need to be precisely $\pm1$. In our simulations we find such a behaviour. For instance, see Fig. \ref{figGab_app} which zooms into Fig. 3(b) of the main text:  $Q$ is only negative at a certain Zeeman strength for which a state of the Andreev band crosses the Fermi level. The value of $Q$ is therefore not a good indicator of the topological phase if non-local conductance is non-zero close to zero energy.  When, instead  $Q$ is calculated in a two-terminal configuration, by removing the right lead, the topological visibility is precisely $Q=+1$ in this case and does not show any dips, since non-local processes are not possible.

\section{Transport calculations}
We perform the transport calculations with the Python package Kwant \cite{Groth2014Kwant}. In particular, we attach normal leads at the left and right end of the nanowire, as schematically shown in Fig.~1 in the main text, inject modes of certain energies and calculate the $S$-matrix. The parameters of the leads are chosen to be the same as in the neighbouring normal section, except the chemical potential, which is set to $\mu_{Lead,L}$ ($\mu_{Lead,R}$) in the left (right) lead. We construct the differential conductance matrix $G_{\alpha \beta}$ with $\alpha, \beta \in \lbrace L,R \rbrace$ from the reflection $r,r'$ and transmission coefficients $t,t'$ of the $S$-matrix
\begin{align}
S=\begin{pmatrix}
r &t\\
t' &r'
\end{pmatrix},
\end{align}
as explained in \cite{Groth2014Kwant,DanonNonlocal2020,Hess2021Local}. 
The transport calculations are performed for different temperatures $T$ (see next section).

\clearpage

\section{System parameters}
Here we provide the parameters used in our calculations. The hyphen in table~\ref{Tab:Parameters} indicates that the corresponding parameter was not used in the calculation.  We note that we used Eq.~\eqref{eq:UniformScGap}  in~\subref*{figDeltaAnd_GFactor_UniformDelta}-\subref*{figNonLocalConductanceAndreevBand_UniformDelta} to model the induced superconducting gap, meaning that there are no interior normal sections.  

\begin{table}[h]
\caption{ Parameter I \label{Tab:Parameters} }
 \centering
\begin{tabular}{c|ccccccccccccccccccccccccc}
\hline
\hline
Fig. & $N_L$ & $N_R$ & $N_S$ &  $N_N$ & $M$& $N_{B,L}$ & $N_{B,R}$&$t_L$ &$t_R$&$t_{SN}$& $\mu_L$ &$\mu_R$ &  $\mu_{SN}$ & $\Delta_0$ & $\Delta_{Z}^c$   \\
\hline
3 & 90 & 94 & 140 & 30 &5  &7 & 7 & 100 \text{meV}& 100 \text{meV}& 20 \text{meV}&0 \text{meV}&0 \text{meV}&2 \text{meV}& 0.25 \text{meV}& 1.75 \text{meV}  \\

2&5&5&150& 10 &6  &5 &5& 158 \text{meV} & 158 \text{meV}&158 \text{meV} & 3 \text{meV}& 3 \text{meV}& 3 \text{meV} & 0.6 \text{meV} & 12.2  \text{meV} \\

\ref{figNonLocalCondDiagram}-\ref{figNonLocalCondDiagram_StandDev}  & 5 & 5 & 150& 5 &-  &5 & 5& 158 \text{meV}&158 \text{meV} & 158 \text{meV}& 3 \text{meV} &3 \text{meV} &3 \text{meV} &0.6 \text{meV} & $\infty$  \\

\ref{figNonLocalCondDiagram_Periodic}-\ref{figNonLocalCondDiagram_ModerateDis}  & 5 & 5 & 150& 5 &-  &5 & 5& 158 \text{meV}&158 \text{meV} & 158 \text{meV}& 3 \text{meV} &3 \text{meV} &3 \text{meV} &0.6 \text{meV} & 21.4    \text{meV}\\

\ref{figDeltaAnd_GFactor_UniformDelta}-\ref{figNonLocalConductanceAndreevBand_UniformDelta}  & 1 & 1 & 28& 2 &6  &1 & 1& 1 & 1& 1&0.3&0.3&0.3& 0.1& 1.265  \\

\ref{figDeltaAnd_GFactor_UniformDelta_NZ_G}-\ref{figNonLocalConductanceAndreevBand_UniformDelta}  & 1 & 1 & 28& 2 &6  &1 & 1& 1 & 1& 1&0.3&0.3&0.3& 0.1& $\infty$  \\

\ref{figAndreevBandDueToVaryingScPhase}    & 1 & 1 & 35& 35 &3  &1 & 1& 1 & 1& 1&0.4&0.4&0.4& 0.1& $\infty$  \\
\hline
\hline
\end{tabular}
\\
\vspace{0.2in}
\begin{tabular}{c|ccccccccccccccccccccccccc}
\hline
\hline
Fig. &  $\alpha_L$ &$\alpha_R$ & $\gamma_L$ & $\gamma_R$& $\mu_{Lead,L}$ &$\mu_{Lead,R}$ & $T$  &$a$  & $g_r$ & $\varphi$\\
\hline
3  &13.75 \text{meV} &13.20 \text{meV} &10 \text{meV}&10 \text{meV}&40 \text{meV} & 40 \text{meV} & 40  \text{mK} & 5 \text{nm} &-&-\\

2 &0  \text{meV} &0  \text{meV} &5  \text{meV}  &5  \text{meV} &5  \text{meV} & 5  \text{meV} & 0  \text{mK} & 4 \text{nm}&-&-\\

\ref{figNonLocalCondDiagram}-\ref{figNonLocalCondDiagram_StandDev}    &0  \text{meV} &0  \text{meV} &4  \text{meV}  &4  \text{meV}  &5  \text{meV}  & 5  \text{meV}  & 50 \text{mK}  & 4 \text{nm}&-&-\\

\ref{figNonLocalCondDiagram_Periodic}-\ref{figNonLocalCondDiagram_ModerateDis}    &0  \text{meV} &0  \text{meV} &4  \text{meV}  &4  \text{meV}  &5  \text{meV}  & 5  \text{meV}  & 50 \text{mK}  & 4 \text{nm}&-&-\\

\ref{figDeltaAnd_GFactor_UniformDelta}-\ref{figNonLocalConductanceAndreevBand_UniformDelta}  &0 &0 &0.25 &0.25 &0.9 & 0.9 & 0  & 1&1&-\\

\ref{figDeltaAnd_GFactor_UniformDelta_NZ_G}-\ref{figNonLocalConductanceAndreevBand_UniformDelta}  &0 &0 &0.25 &0.25 &0.9 & 0.9 & 0  & 1&0.8&-\\

\ref{figAndreevBandDueToVaryingScPhase}  &0 &0 &0.25 &0.25 &0.9 & 0.9 & 0  & 1& -&$\pi/1.4$\\

\hline
\hline
\end{tabular}
\end{table}

\begin{table}[h]
\caption{ Parameter II  \label{Tab:ParametersTopologicalWire} }
 \centering
\begin{tabular}{c|cccccccccccccccccccc}
\hline
\hline
Fig. & $N_L$ & $N_R$ & $N_S$ &  $N_N$ & $M$& $N_{B,L}$ & $N_{B,R}$&$t$& $\mu_L$ &  $\mu_{SN}$ & $\Delta_0$ &   $\kappa_1$ &   $\kappa_2$ & $a$& $\lambda$ & $T$ & $\alpha$ & $\gamma_L=\gamma_R$& $\mu_{Lead,L}=\mu_{Lead,R}$\\
\hline
\ref{figGab_app}  & 11 & 1 & 151& 0 &1  &1 & 1& 1 & 0&0.2&0.1 & 0.95 & 0.015  &1 & 5 & 0 & 0.2 & 0.5 & 0.6\\
\hline
\hline
\end{tabular}
\end{table}

\vspace{0.5in}

\bibliographystyle{apsrev4-1}
\bibliography{Literatur2}